\documentclass[prl,preprintnumbers,amsmath,amssymb,twocolumn,superscriptaddress]{revtex4}

\usepackage{amsfonts}
\usepackage{amssymb}
\usepackage{mathbbol}
\usepackage{amsfonts}
\usepackage{mathrsfs}
\usepackage{epsfig,bm,dcolumn}
\usepackage{graphicx}
\usepackage{color}
\usepackage{amsmath}
\usepackage{overpic}
\usepackage{slashed}
\usepackage{hyperref}
\usepackage{float}
\usepackage{varwidth}
\usepackage{tabularx}
\usepackage{multirow}

\usepackage[nottoc]{tocbibind}
\usepackage{framed} 
\begin{document}

\title{Continuous-mixture Autoregressive Networks for efficient variational calculation of many-body systems}
\author{Lingxiao Wang\footnote{lwang@fias.uni-frankfurt.de}}
\affiliation{Frankfurt Institute for Advanced Studies, Ruth-Moufang-Str. 1, 60438 Frankfurt am Main, Germany}
\affiliation{Department of Physics, Tsinghua University, Beijing 100084, China.}

\author{Yin Jiang\footnote{jiang\_y@buaa.edu.cn}}
\affiliation{Department of Physics, Beihang University, Beijing 100191, China.}
\author{Lianyi He\footnote{lianyi@mail.tsinghua.edu.cn}}
\affiliation{Department of Physics, Tsinghua University, Beijing 100084, China.}
\author{Kai Zhou\footnote{zhou@fias.uni-frankfurt.de}}
\affiliation{Frankfurt Institute for Advanced Studies, Ruth-Moufang-Str. 1, 60438 Frankfurt am Main, Germany}

\date{\today}

\begin{abstract}
We develop deep autoregressive networks with multi channels to compute many-body systems with \emph{continuous} spin degrees of freedom directly. As a concrete example, we embed the two-dimensional XY model into the continuous-mixture networks and rediscover the Kosterlitz-Thouless (KT) phase transition on a periodic square lattice. Vortices characterizing the quasi-long range order  are accurately detected by the autoregressive neural networks. By learning the microscopic probability distributions from the macroscopic thermal distribution, the neural networks compute the free energy directly and find that free vortices and anti-vortices emerge in the high-temperature regime. As a more precise evaluation, we compute the helicity modulus to determine the KT transition temperature. Although the training process becomes more time-consuming with larger lattice sizes, the training time remains unchanged around the KT transition temperature. The continuous-mixture autoregressive networks we developed thus can be potentially used to study other many-body systems with continuous degrees of freedom.
\end{abstract}

\maketitle

\textit{Introduction}.---Machine learning techniques are attracting widespread interest in different fields of science and technology because of its power to extract and express structures inside complex data. In particular, physicists have employed it to do projects including classification, regression, and pattern generation~\cite{buchanan:2019power, carleo:2019machine}. It was found that the neural networks can classify the phase structures of many-body systems ~\cite{wang:2016discovering,carrasquilla:2017machine,pang:2018equationofstatemeter,fujimoto:2018methodology,fujimoto:2020mapping}. As for the regression, it was successfully applied to the event selection in a large data set (e.g., from the LHCb)~\cite{metodiev:2018jet,kasieczka:2019machine}, the spinodal decomposition in heavy ion collisions~\cite{steinheimer:2019machine}, and the molecular structure prediction~\cite{smith:2017ani1}. Furthermore, it also sheds light on the innovation of the methods of first-principle calculations of many-body systems~\cite{carleo:2017solving,nagy:2019variational, hartmann:2019neuralnetwork,pfau:2019abinitio, vicentini:2019variational, yoshioka:2019constructing}. The restricted Boltzmann machine was applied to solve quantum many-body systems~\cite{carleo:2017solving}. A deep neural network was constructed to derive the solution of the many-electron schr\"odinger equation~\cite{pfau:2019abinitio}. It was found that proper neural networks can work as an efficient $Ansatz$ to characterize many-body systems. Another interesting direction is modification of the classical algorithms with machine learning ~\cite{shen:2018selflearning,mori:2018application,carleo:2019machine}, which may improve or assist the conventional first-principle computations. 

In addition, it is natural to apply neural networks to many-body systems on the lattices, since they share similar discrete architectures. In some pioneer attempts~\cite{alexandru:2017deep, broecker:2017machine, urban:2018reducing, mori:2018application,zhou:2019regressive}, the training data were generated by the classical Markov Chain Monte Carlo (MCMC) method. However,  its expandability and efficiency are limited because of the critical slowing down near the critical point~\cite{urban:2018reducing} and the sign-problem~\cite{broecker:2017machine}. Recently, a new method based on an autoregressive neural network was proposed and applied to discrete spin systems (e.g., the Ising model)~\cite{wu:2019solving,sharir:2020deep}. This new method uses a variational \textit{Ansatz} that decomposes the macroscopic probability distribution 
into microscopic distributions on the lattice sites. It was demonstrated that a higher computational accuracy can be achieved in solving several Ising-type systems~\cite{ou:2019review}.

However, it still remains challenging to solve a general many-body system with continuous degrees of freedom, e.g., a continuous spin system which may exhibit a topological phase transition. 
Previous attempts failed when one applied methods that work well for discrete spin systems to continuous cases~\cite{cristoforetti:2017meaningful}.
Being different from the classical phase transition, the topological phase transition occurs with topological defects emerging. This has been attracting the attention from various fields of physics~\cite{scientificbackground:2016topological}. Some related efforts using both supervised learning and unsupervised learning have been made to handle this problem~\cite{wang:2017machine,beach:2018machine,suchsland:2018parameter,zhang:2018machine,carvalho:2018realspace,hu:2019machine,fukushima:2019featuring}. The winding numbers were recognized by a neural network with supervised training for an one-dimensional insulator model~\cite{zhang:2018machine}. By generating the configurations with MCMC sampling and supplying feature engineered vortex configurations as the input, neural networks could detect the topological phase transition from well-preprocessed configurations~\cite{beach:2018machine}. While the unsupervised learning was applied to identify the topological orders~\cite{cristoforetti:2017meaningful,rodriguez-nieva:2019identifying,scheurer:2020unsupervised}, an efficient variational approach combined with powerful autoregressive neural networks is still missing.

In this Letter, we propose Continuous-mixture Autoregressive Networks (CANs) to solve the continuous spin systems efficiently, which can be further applied to continuous field systems. As a concise reference example, we study the two-dimensional (2D) XY model on a square lattice, which exhibits the so-called Kosterlitz-Thouless (KT) phase transition~\cite{gupta:1988phase,kosterlitz:1974critical,weber:1988monte}. The CANs are introduced to recognize the topological phase transition with continuous variables in an unsupervised manner. In such autoregressive neural networks, the microscopic state at each lattice site is modeled by a conditional probability, which constructs a joint probability for the whole configuration~\cite{wu:2019solving,sharir:2020deep}.  In this work, we introduce a generic framework for CANs and construct suitable neural networks to study the XY model. The vortices, serving as the signal of the KT transition, are automatically generated by the neural networks. Correspondingly, the KT transition temperature of the 2D XY model can be more accurately determined by calculating the helicity modulus~\cite{weber:1988monte}. As for the computing time cost, the training process becomes undoubtedly more time-consuming for larger lattice sizes. However, the training time remains unchanged around the transition point. Considering the advantages of the CANs, we propose further potential applications to other many-body systems with continuous degrees of freedom in the final part of this work.

\begin{figure}[hpt!]
\centering
	\includegraphics[width=8cm]{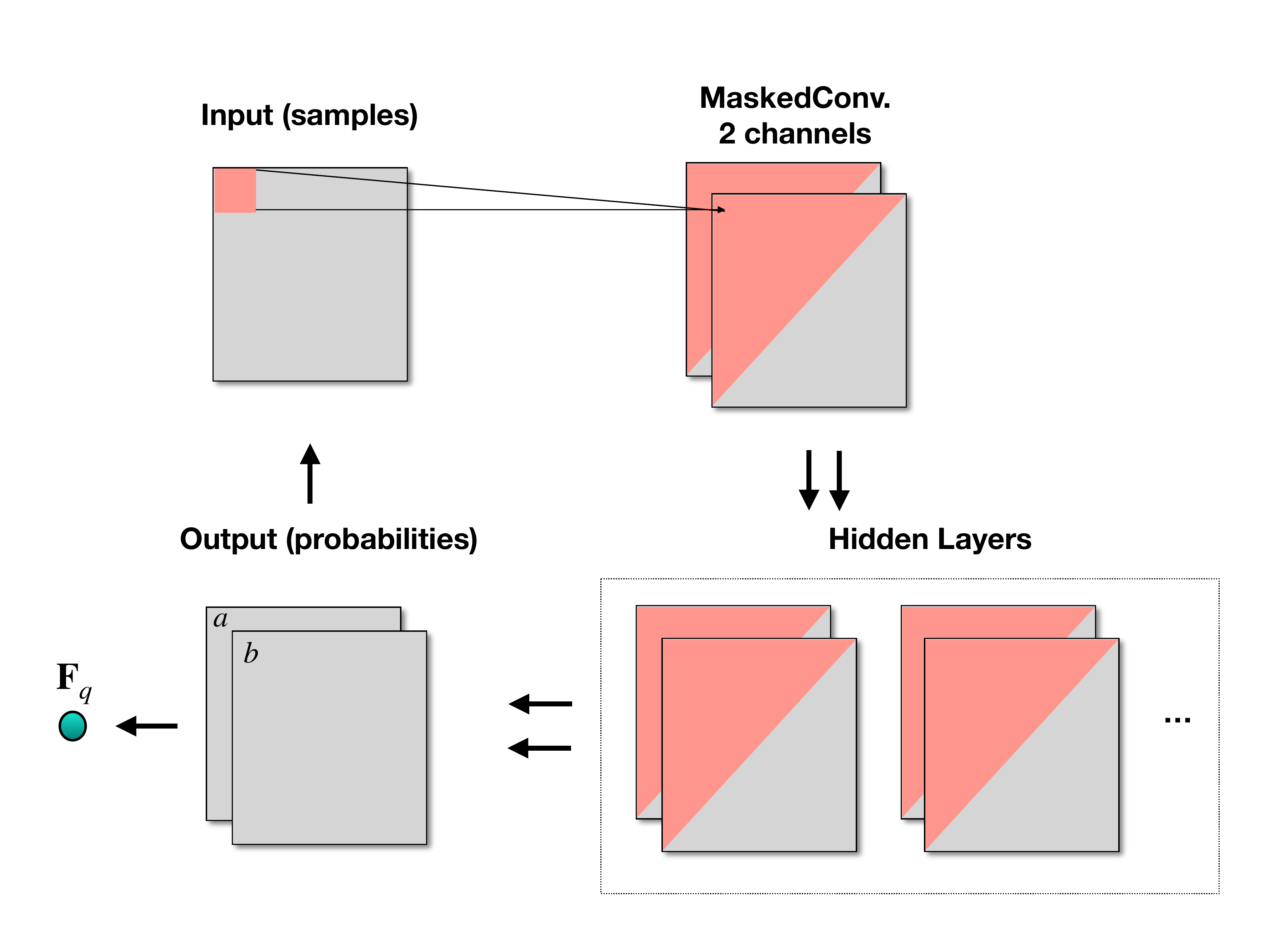}
	\caption{The architecture of the Continuous-mixture Autoregressive Networks(CANs) for the computation of a continuous spin system. It 
	is a continuous-mixture PxielCNN structure~\cite{vandenoord:2016pixel}, in which the masked layers are added into the network to establish the autoregressive networks.}
	\label{fig:cans}
\end{figure}

\textit{Continuous-mixture Autoregressive Networks}\label{sec:can}.---Let us consider a many-body system on a lattice with continuous spin degrees of freedom at each lattice site. Because the orientation of each spin changes continuously, the conditional probability $q(s_i)$ for spin $s_i$ at site $i$ must also be continuous. We propose that a proper mixture of the beta distribution ${\rm Beta}(a, b)$ is the prior probability to ensure that the continuous variables distribute randomly in a finite interval~\cite{betadis}. The beta distribution ${\rm Beta}(a, b)$ is continuously defined in a finite interval with two positive shape parameters $a, b$. Thus the output layers of neural networks are designed to be of two channels for each Beta component, and the conditional probabilities are derived as
\begin{equation}
	f_\theta(s_i|s_1,...,s_{i-1})=\frac{\Gamma(a_i+b_i)}{\Gamma(a_i) \Gamma(b_i)} s_i^{a_i-1}(1-s_i)^{b_i-1},
\end{equation}
where $\Gamma(a)$ is the gamma function, $\{\theta\}$ is a set of parameters of the networks, and $s_i=\phi_i/2\pi\in[0,1]$. The outputs of the hidden layers are $\mathbf{a}\equiv(a_1,a_2,\cdots)$ and $\mathbf{b}\equiv(b_1,b_2,\cdots)$, which can be realized by Fig.~\ref{fig:cans} as an autoregressive neural networks.

%%%%%%%%%%%%%%%%%%%%%%%%%%%%%%%%%%%%%%%%%%%%%%%%%%%%%
%%%%%%%%%%%%%%%%%%%%%%%%%%%%%%%%%%%%%%%%%%%%%%%%%%%%%

As a practical example, we consider the 2D XY model on the lattice. The Hamiltonian is expressed in terms of spins living at the lattice sites with nearest-neighbor interactions,
\begin{equation}
	H=-J\sum_{<i,j>}s_is_j=-J\sum_{<i,j>}\cos(\phi_i-\phi_j),
\end{equation}
 where $<i,j>$ indicates that the sum is taken over all nearest-neighbor pairs and the angle $\phi_i\in[0,2\pi)$ denotes the spin orientation at site $i$.  The Mermin-Wagner theorem forbids long-range order (LRO) in 2D systems with continuous degrees of freedom, since the strong fluctuations in 2D break the order~\cite{wagner:2010merminwagner}. Nevertheless, the formation of topological defects (i.e., vortices and anti-vortices) in the XY model distinguishes phases with and without quasi-LRO, which characterizes the global properties of the many-body system.

 To detect the KT phase transition in the XY model, we study the free energy $F=-(1/\beta)\ln Z$ according to statistical mechanics, where $\beta\equiv 1/(k_BT)$, with $T$ being the temperature. The free energy is obtained from the partition function $Z\equiv \sum_{\mathbf{s}}\exp(-\beta E(\mathbf{s}))$, which contains all information of the system. The summation runs over all possible configurations $\{\mathbf{s}\}$ of the system. Monte Carlo algorithms can be routinely applied to generate the configurations and can achieve proper relative importance among different configurations. However, the free energy cannot be computed directly from the Monte Carlo algorithms. Variational approaches can be employed to obtain a variational free energy. In this work, the variational target function is set to be the joint probability of all configurations and equals the Boltzmann distribution $p(\mathbf{s})={e^{-\beta E(\textbf{s})}}/{Z}$.
 %The configurations $\mathbf{s}=\{s_1,s_2,...,s_N\}$ with continuous spins are defined on the lattice with $N$ sites.
 We use a variational ${Ansatz}$ for the joint distribution, denoted by $q_\theta(\mathbf{s})$. It is parametrized by a set of variational parameters $\{\theta\}$, which are tuned to approach the target distribution $p(\mathbf{s})$. The Kullback-Leibler divergence~\cite{mackay:2003information} between the variational and the target distributions, $D_{\mathrm{KL}}\left(q_{\theta} \| p\right)\equiv \mathbb{E}_{\mathbf{s} \sim q_{\theta} }(-\log p+\log q_{\theta})$, provides the measure of the closeness from $q_\theta$ to $p$. The corresponding variational free energy $F_{q}$ can be derived from $ D_{\mathrm{KL}}\left(q_{\theta} \| p\right)=\beta\left(F_{q}-F\right)$. We obtain  
 \begin{equation}
 	F_{q}=\frac{1}{\beta}\sum_{\mathbf{s}} q_{\theta}(\mathbf{s})\left[\beta E(\mathbf{s})+\ln q_{\theta}(\mathbf{s})\right].
 	\label{eq:f}
 \end{equation}
 Since $ D_{\mathrm{KL}}\left(q_{\theta} \| p\right)$ is non-negative, the variational free energy $F_{q}$ gives an upper bound of the true free energy $F$. Thus the minimization of $ D_{\mathrm{KL}}\left(q_{\theta} \| p\right)$ and the variational free energy $F_{q}$ are actually equivalent. Meanwhile, as pointed out in previous works~\cite{carleo:2017solving,wu:2019solving,sharir:2020deep}, it is straightforward to map the parameters $\{\theta\}$ onto the weights of an Artificial Neural Network (ANN). Correspondingly, the variational free energy becomes the loss function. Using the log-derivative trick~\cite{williams:1992simple} we obtain
\begin{equation}
\beta \nabla_{\theta} F_{q}=\mathbb{E}_{\mathbf{s} \sim q_{\theta}}\Big\{\left[\beta E(\mathbf{s})+\ln q_{\theta}(\mathbf{s})\right] \nabla_{\theta} \ln q_{\theta}(\mathbf{s})\Big\},
	\label{gradient}
\end{equation}
where the gradient $\nabla_{\theta} \ln q_{\theta}(\mathbf{s})$ is weighted by the reward signal $\beta E(\mathbf{s})+\ln q_{\theta}(\mathbf{s})$.

Once the parameters $\{\theta\}$ are mapped onto ANN, the variational problem becomes nothing but training the networks. Using autoregressive networks such as CANs for the parametrization, we can decompose the variational distribution into a product of the conditional probabilities,
\begin{equation}
	q_{\theta}(\mathbf{s})=\prod_{i=1}^{N} f_{\theta}\left(s_{i} | s_{1}, \ldots, s_{i-1}\right)\label{ansatz},
\end{equation}
which provides the variational ${Ansatz}$ by parametrizing each conditional probability as neural networks. Considering the symmetry, we employ the PixelCNN\cite{vandenoord:2016pixel} that can naturally preserve the locality and the translational symmetry on a square lattice. In addition, the autoregressive property is guaranteed by putting a mask on the convolution kernel~\cite{germain:2015made,maskcnn}. As shown in Fig.~\ref{fig:cans}, the input layer takes in the configurations $\mathbf{s}$ on the lattice. After passing it through several masked convolution layers, the parameters of the beta distribution at each site are obtained in the output layer. Then the configuration probability $q_{\theta}({\bf s})$ can be derived and the variational free energy can be further calculated via Eq.\eqref{eq:f} after such a forward propagation for a batch of independent configurations. If we specify the channels in the convolution layers to represent the parameters of each beta component, we find that it greatly saves the training time and speeds up the sampling.  

Training the neural networks here is the key to perform the variational approach. We use a classical back-propagation algorithm. The nuts-and-bolts computation with CANs is with the following procedures: (i) With the randomly initialized network, sample a batch of independent configurations to be the training set; (ii) Pass the training set forward to evaluate the log-probability $\ln q_{\theta}$ and the variational free energy $F_q$; (iii) Estimate the gradient $\nabla_{\theta} F_{q}$ and update the network weights via back-propagation; (iv) With the updated network, resample a batch of configurations to be the new training set, which actually follow the current joint probability $q_\theta$; (v) Repeat the above procedures until the loss function becomes eventually convergent; (vi) Sample an ensemble of independent configurations from $q_\theta$ site by site at once; (vii) Calculate the thermodynamic observables. 
On the square lattice with $N$ sites, the convergence is reached if the change of the variational energy per site is much smaller than the superior limit, $\delta F_q\ll4J/N$~\cite{chung:1999essential}.

\textit{Rediscovery of KT Transition}\label{sec:results}.---Now we study the KT transition and calculate the thermodynamic quantities of the 2D XY model using CANs. In the calculations, the default width and depth of the network we adopted in CANs are set to be $(32, 3)$, with $N=L^2$ lattice sites being the input. The multi-channel feature is used to construct a mixture of the beta distributions, which makes the networks more expressive~\cite{salimans:2017pixelcnn}. The Adam optimizer is applied to minimize the loss function in Pytorch. The CANs can be implemented on GitHub~\cite{wang:2020can2xy} and the corresponding hyper-parameters are consistent with previous works~\cite{wu:2019solving}. 

%%%%%%%%%%%%%%%%%%%%%%%%%%%%%%%%%%%%%%%%%%%%%%%%%%%%%
\begin{figure}[H]
	\centering
	\includegraphics[width=7cm]{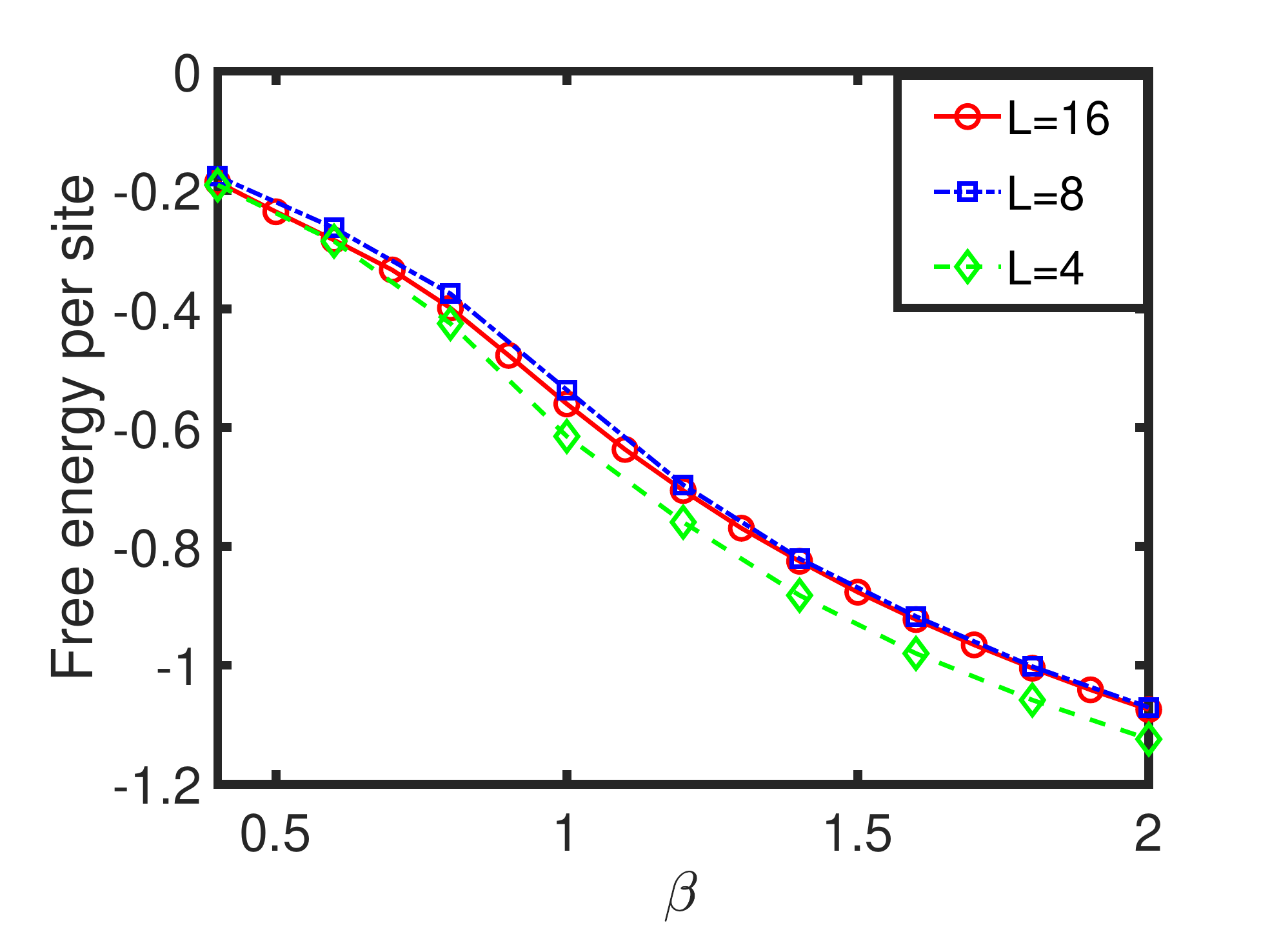}
	\caption{The free energy per site of the 2D XY model on a square lattice from CANs.}
	\label{fig:fenergy}
\end{figure}
%%%%%%%%%%%%%%%%%%%%%%%%%%%%%%%%%%%%%%%%%%%%%%%%%%%%%
%%%%%%%%%%%%%%%%%%%%%%%%%%%%%%%%%%%%%%%%%%%%%%%%%%%%%
\begin{figure}[H]
	\centering
	\includegraphics[width=8cm]{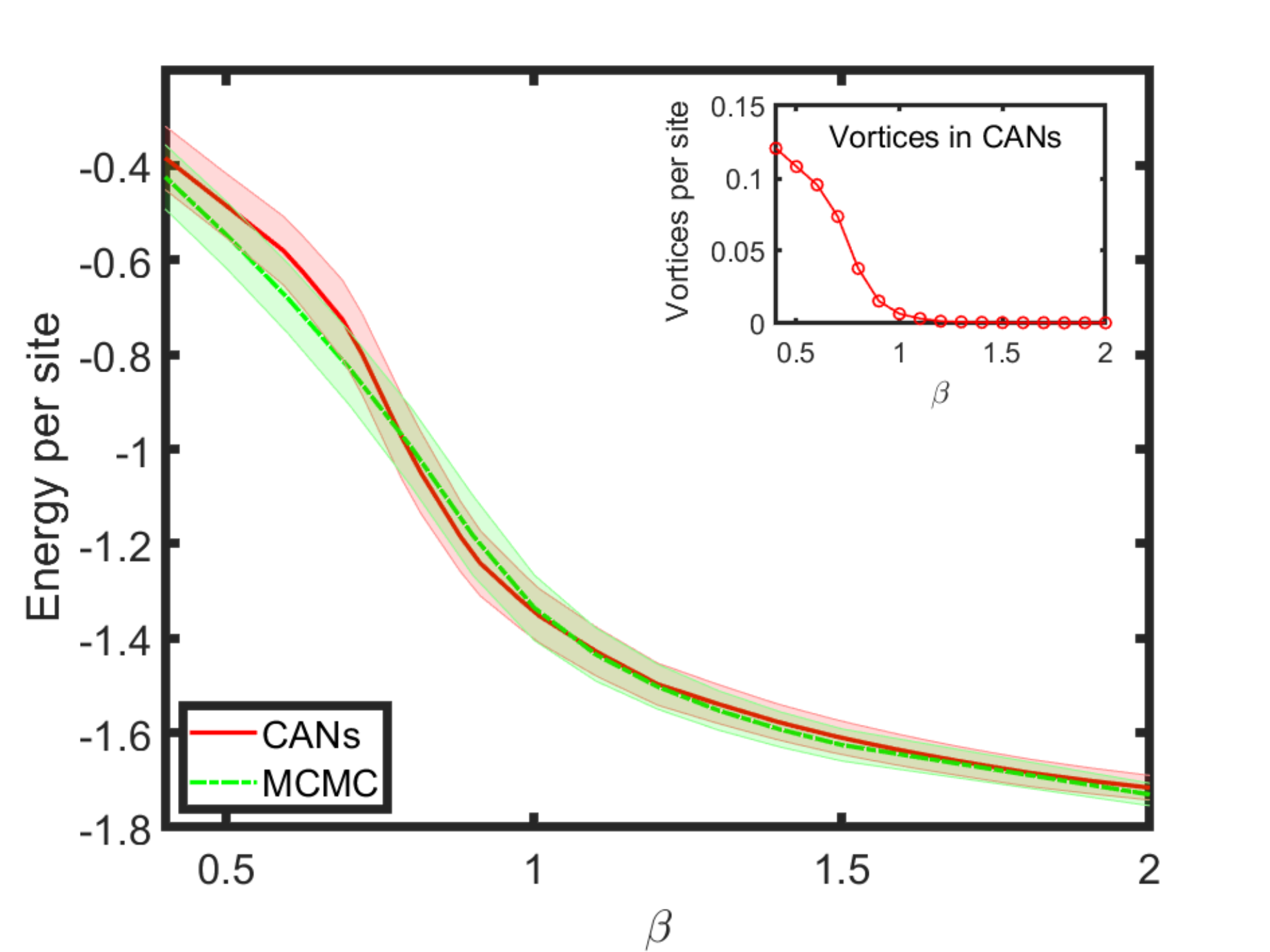}
	\caption{The energy per site as a function of $\beta$ with lattice size $L=16$ from CANs with a 100-channel mixture of the Beta distribution. The shaded area denotes the statistical error from thermal fluctuations. As a comparison, we also show the result from MCMC. The insert shows the number of vortex pairs per site extracted from CANs.}
	\label{fig:comparison}
\end{figure}
%%%%%%%%%%%%%%%%%%%%%%%%%%%%%%%%%%%%%%%%%%%%%%%%%%%%%

The thermodynamic observables of the 2D XY model have been computed with the MCMC method~\cite{gupta:1988phase,hasenbusch:2005twodimensionalxymodel,komura:2012largescale,weber:1988monte}. It is interesting to compare the results from CANs and MCMC. The advantage of CANs is that the free energy per site can be directly calculated. It is presented Fig.~ \ref{fig:fenergy} for three different lattice sizes,  $L=4, 8, 16$. The results for $L=8$ and $L=16$ indicate that the free energy converges rapidly with increasing lattice sizes, which ensures that the size effect can be avoided.  In the following discussions we use the results for $L=16$. 

Fig.~\ref{fig:comparison} shows the energy per site as a function of $\beta$ obtained from CANs and MCMC with the same number of configurations. Here the MCMC calculation was implemented with a classical algorithm~\cite{tobochnik:1979monte,teitel:1983phase}. In the low temperature regime ($\beta>1$), the result from CANs agrees with that from MCMC despite the statistical error from thermal fluctuations. In the high temperature regime ($\beta<1$), however, the two results do not match perfectly.  We can understand this deviation from the comparison of the  number of free vortices (or anti-vortices) shown in Table~\ref{tab:vor}. Since a larger number of free vortices and anti-vortices indicates a larger entropy in the XY model, configurations with more vorticity~\cite{vorticity} lead to a lower free energy. This is the reason why the energy per site from CANs is larger than that from MCMC at $\beta<0.7$, because the entropy from the free vortices and anti-vortices balances the free energy. The situation is reversed at $0.7<\beta<1$ for the same reason. The rapid increase of the number of free vortices per site around $\beta\sim 1$ indicates that there exists a topological phase transition, i.e., the KT transition.
%%%%%%%%%%%%%%%%%%%%%%%%%%%%%%%%%%%%%%%%%%%%%%%%%%%%%
\begin{table}[H]
\caption{The energy and the number of free vortices (or anti-vortices) per site extracted from CANs and MCMC. The results from CANs are obtained from an ensemble average with 1000 configurations.}
\label{tab:vor}
\centering
\begin{tabular}{ccccccc}
\\
\hline \hline
\multicolumn{2}{c}{\textbf{$\beta$}}  & 0.4 & 0.6 & 0.8 & 1.0 & 1.2 \\
 \hline
\multirow{2}{*}{\textbf{Energy}}& MCMC & -0.424   & -0.682   & -0.996   & -1.336   & -1.502   \\& CANs & -0.384   & -0.588   & -1.017   & -1.346   & -1.497 \\
\hline
\multirow{2}{*}{\textbf{Vortices}} & MCMC & 0.114   & 0.080   & 0.042   & 0.010   & 0.002   \\& CANs & 0.121  & 0.096 & 0.038   & 0.007   & 0.001  \\
\hline \hline
\end{tabular}
\end{table}
%%%%%%%%%%%%%%%%%%%%%%%%%%%%%%%%%%%%%%%%%%%%%%%%%%%%%
The KT transition is a transition from bound vortex-antivortex pairs at low temperatures to unpaired vortices and anti-vortices at high temperature. In previous works~\cite{gupta:1988phase, komura:2012largescale, bighin:2019berezinskiikosterlitzthouless}, the KT transition temperature was reported as $\beta_{\rm KT}\approx 1.12$. In our method, the CANs capture the global property of the 2D XY model since the trained neural networks help achieve a good evaluation of the free energy. Nevertheless, at high temperature, the disorder of the configurations due to the thermal fluctuations results in a slight mismatch between CANs and MCMC. The temperature effect attenuates the long-range correlation exponentially~\cite{kosterlitz:1974critical}, which also slightly weakens the expressive ability of CANs for a finite-size system. Additionally, we emphasize that the elementary conditional distribution at each site, $f_{\theta}\left(s_{i} | s_{1}, \ldots, s_{i-1}\right)$, should be carefully chosen since the spin in the XY model is periodically valued. We find that, as a common test, the choice of the normal distribution for $f_{\theta}\left(s_{i} | s_{1}, \ldots, s_{i-1}\right)$ can hardly make the the loss function converge to a reasonable value.

\textit{KT Transition Point}.---To recognize the KT transition point quantitatively, we may introduce the spin stiffness $\rho_s$, which reflects the change of the free energy in response to an 
infinitesimally slow twist $\delta\phi$ on the spins. In the continuum limit, it is $\rho_s=[{\partial^2 F(\delta \phi)}/{\partial (\delta \phi)^2 }]|_{\delta \phi=0}$. In practice, we consider another quantity, the helicity modulus $\gamma(L)$~\cite{weber:1988monte,gupta:1988phase,komura:2012largescale,hasenbusch:2005twodimensionalxymodel}, which is equivalent to $\rho_s$ in the limit $\delta\phi\rightarrow0$. It can be expressed as
\begin{equation}
	\gamma(L) = -\frac{E}{2 L^2} - \frac{J \beta}{L^2} \left\langle\Big(\sum_{<i,j>}\sin(\phi_i-\phi_j){\bf e}_{ij}\cdot{\bf x}\Big)^2\right\rangle,
\end{equation}
where $L$ is the size of the square lattice, ${\bf e}_{ij}$ is the vector pointing from site $j$ to site $i$, and ${\bf x}$ is an arbitrary unit vector in the 2D plane. The Kosterlitz renormalization group~\cite{kosterlitz:1974critical} predicts that $\gamma(L\to\infty )$ jumps from the value $2 k_BT_ c /\pi$ to zero at the critical temperature, and hence the helicity modulus gives a reliable prediction of the KT transition point.
%%%%%%%%%%%%%%%%%%%%%%%%%%%%%%%%%%%%%%%%%%%%%%%%%%%%%
\begin{figure}[H]
	\centering
	\includegraphics[width=8cm]{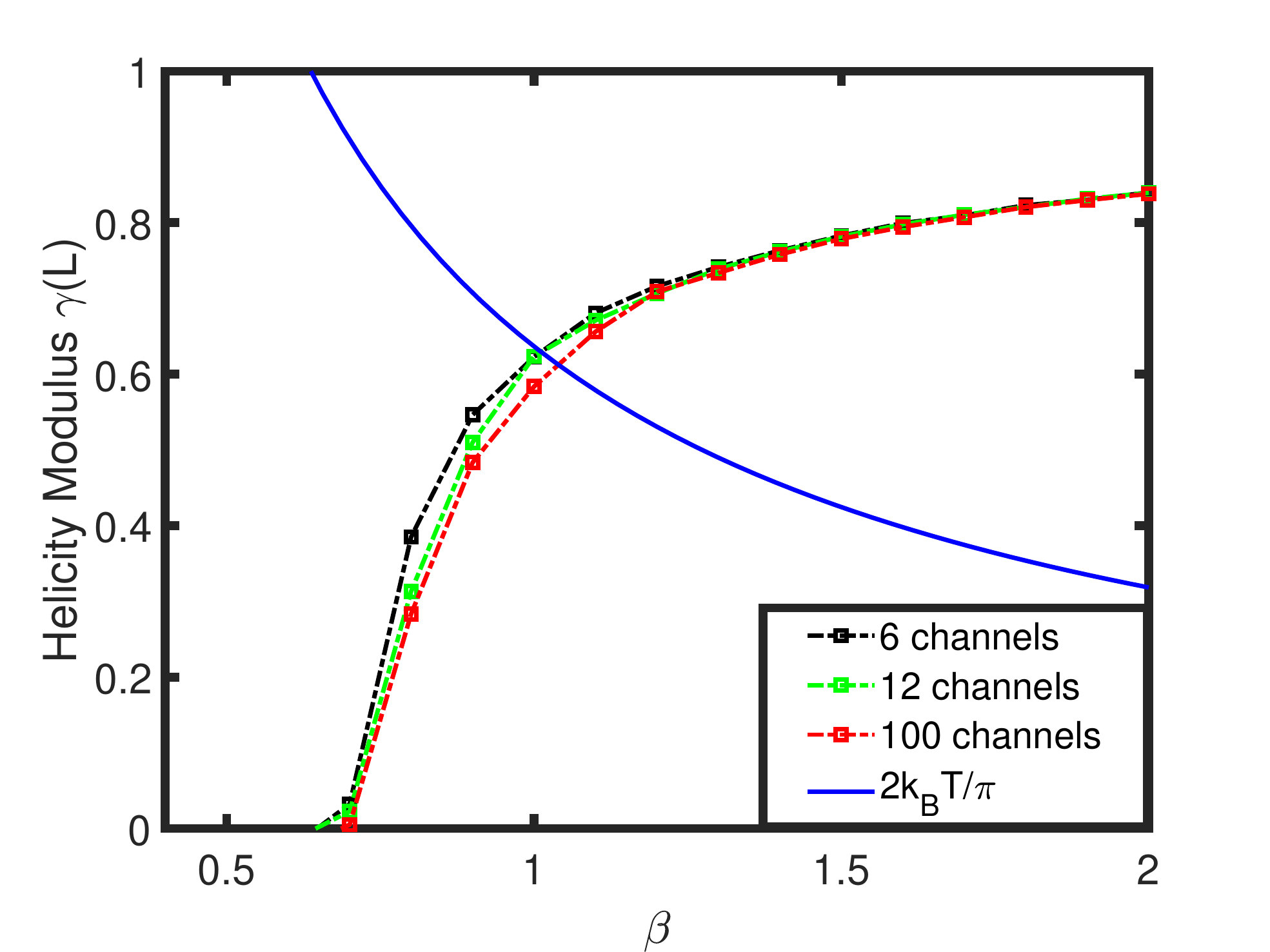}
	\caption{The helicity modulus as a function of $\beta$ for lattice size $L=16$ obtained from CANs with a 100-channel mixture of the beta distribution. The cross point with the curve $2k_BT/\pi$ gives the transition point $\beta_{\rm KT}\simeq 1.10$.}
	\label{fig:criticalpoint}
\end{figure}
%%%%%%%%%%%%%%%%%%%%%%%%%%%%%%%%%%%%%%%%%%%%%%%%%%%%%

In Fig.~\ref{fig:criticalpoint}, we show the helicity modulus for a large lattice size $L=16$. The markers are the numerical results evaluated from CANs with several multi-channel mixtures of the beta distribution and the dashed lines are the corresponding interpolation curves. With increasing number of channels, the crossing point with the curve $2 k_BT/\pi$ predicts the KT transition temperature. For a sufficiently large number of channels (100), we observe that the crossing point is located at $\beta\simeq 1.10$. Since the helicity modulus depends on the correlation function that needs a higher order statistics than the energy (i.e., the mean square of the energy), here we only consider a large lattice size that can give a precise evaluation of $\gamma$. The result is consistent with that from the standard Monte Carlo simulations ~\cite{komura:2012largescale,beach:2018machine}. 

We find that the computational time cost is around 0.2 seconds per training step and is almost independent of the temperature. This indicates that the Critical Slowing Down (CSD) is hopefully avoided. Even though the burden due to the increasing auto-correlation time~\cite{goodman:1989multigrid,urban:2018reducing} in MCMC does not appear in CANs, the relation between the training time cost and the lattice size should be mentioned here. The training time per step in CANs with the default network setup for different lattice sizes at the KT transition temperature can be recorded. We find that as a function of the lattice size $L$ it can be well fitted as $t_{\text{train}}(L)=a\,L^b$, with $a = 0.0026$ and $b = 1.708$. Thus the cost of bypassing the CSD is such a training time that has an approximate polynomial dependence on $L$. However, it is just one-off and the following configuration sampling procedure can take on the parallel advantage of GPU for a large ensemble generation. Therefore, a more powerful GPU can reduce the time cost efficiently. As a reference, all results presented in this work were obtained on a Nvidia RTX 2080 GPU.

\textit{Summary}.---We have proposed continuous-mixture autoregressive networks (CANs) for an efficient variational calculation of many-body systems with continuous degrees of freedom. Specifically, a CAN can be designed as an \textit{Ansatz} for the variational approach to the topological phase transition in the 2D XY model. The CANs are able to learn to construct microscopic states of the system, in which vortices and anti-vortices emerge automatically. We can evaluate the energy and the vorticity using configurations generated from the networks and compare them to the results from the MCMC calculations. The autoregressive structure of the neural networks is beneficial to study the long-range correlations even beyond the phase transition point. It sheds light on the investigation of more latent topological structures, such as a coupled XY model~\cite{bighin:2019berezinskiikosterlitzthouless} and a twisted bilayer graphene~\cite{julku:2020superfluid}, where novel long-range correlations may emerge.  Besides, a straightforward determination of the KT transition point in CANs is shown to be consistent with previous works using the standard Monte Carlo methods. 

Although the increasing time cost with the lattice size is unavoidable,  it becomes more economical in searching for the critical point in the limit of a large ensemble of configurations, because of the direct GPU usage. With the help of CANs, the CSD problem occurring in MCMC is expected to be remarkably alleviated, which may
reduce the computational difficulties in more complicated many-body systems, e.g., lattice simulation for the possible critical end point in the QCD phase diagram. The $\phi^4$ model could be a practical step~\cite{urban:2018reducing}, where the computational accuracy and practicability could be rigorously examined. Therefore, the deep learning approach, especially with well-designed neural networks, can be applied to specific physical problems~\cite{mehta:2014exact,iten:2020discovering}. This inspires us to explore the techniques from a more physical viewpoint, which will help us open the black boxes of the deep learning and the nature.

~\
\begin{acknowledgments}
We thank Giuseppe Carleo and Junwei Liu for useful discussions. The work is supported by the BMBF under the ErUM-Data project (K. Z.), the AI grant of SAMSON AG, Frankfurt (K. Z.),  the National Natural Science Foundation of China with Grant No. 11875002 (Y. J.) and No.11775123 (L. H. and L. W.), the Zhuobai Program of Beihang University (Y. J.), and the National Key R\&D Program of China with Grant No. 2018YFA0306503 (L. H.). K. Z. also thanks the donation of NVIDIA GPUs from NVIDIA Corporation.
\end{acknowledgments}

\bibliography{CANsXY.bib}

\begin{thebibliography}{60}
\bibitem [{bet()}]{betadis}%
  \BibitemOpen
  \href@noop {} {}\bibinfo {note} {Intuitively, in the Bayesian inference the
  beta distribution is the conjugate prior probability distribution of the
  Bernoulli distribution, which has been proven to be a proper distribution for
  the Ising model case.}\BibitemShut {Stop}%
\bibitem [{mas()}]{maskcnn}%
  \BibitemOpen
  \href@noop {} {}\bibinfo {note} {This also ensures that the weights are
  nonzero for half of the kernel and each conditional probability
  $f_{\theta}(s_i|s_1,...,s_{i-1})$ is independent of $s_j$ with $j>i$ for a
  prechosen ordering.}\BibitemShut {Stop}%
\bibitem [{vor()}]{vorticity}%
  \BibitemOpen
  \href@noop {} {}\bibinfo {note} {The number of free vortices (which equals to
  the number of free anti-vortices) is given by $n = v/(2 L^2)$, where the
  vorticity is given by $v = (1/2\pi) \oint_C \nabla \phi(\boldsymbol{r})\cdot
  d\boldsymbol{r}$ in the continuum limit.}\BibitemShut {Stop}%
\end{thebibliography}
\end{document}